\documentclass[preprint2]{aastex63}
\pdfoutput=1
\usepackage{savesym}
\savesymbol{tablenum}
\usepackage{siunitx}
\restoresymbol{SIX}{tablenum}
\usepackage[utf8]{inputenc}
\usepackage{array,appendix, bm, amsmath, amsfonts, esint, bbold, graphicx,appendix,tabu, float, siunitx,hyperref, gensymb, comment, geometry, times, epstopdf, lineno}
\usepackage[caption=false]{subfig}

\DeclareUnicodeCharacter{2212}{-}
\makeatletter
\DeclareRobustCommand{\HI}{%
 \mbox{H\check@mathfonts\fontsize \sf@size\z@\selectfont I }%
}
\usepackage{hyphenat}
\shorttitle{GALFA-\HI SPS}
\shortauthors{Mittal  et al.}
\title{GALFA-SPS}
\date{July 2023}
\begin{document}

\title{Neutral Hydrogen (\HI) 21 cm as a probe: Investigating Spatial Variations in Interstellar Turbulent Properties}

\author{Amit Kumar Mittal}
\affiliation{
Department of Astronomy, University of Wisconsin-Madison,
475 N Charter Street, Madison, WI, 53706-1582 USA}
\affiliation{Department Electrical and Computer Engineering, University of Wisconsin-Madison,
1415 Engineering Dr, Madison, WI, 53706-1582 USA}

\author{Brian L Babler}
\affiliation{
Department of Astronomy, University of Wisconsin-Madison,
475 N Charter Street, Madison, WI, 53706-1582 USA}

\author[0000-0002-3418-7817]
{Sne{\v{z}ana} Stanimirovi{\'c}}
\affiliation{
Department of Astronomy, University of Wisconsin-Madison,
475 N Charter Street, Madison, WI, 53706-1582 USA}

\author{Nickolas Pingel}
\affiliation{
Department of Astronomy, University of Wisconsin-Madison,
475 N Charter Street, Madison, WI, 53706-1582 USA}


\begin{abstract}
 Interstellar turbulence shapes the \HI\ distribution in the Milky Way (MW). How this affects large-scale statistical properties of \HI\ column density across the MW remains largely unconstrained.
We use the $\sim13,000$ square-degree GALFA-\HI\ survey
to map statistical fluctuations of \HI\
over the $\pm 40$ km s$^{-1}$ velocity range.
We calculate the spatial power spectrum (SPS) of \HI\ column density image by running a 3-degree kernel and measuring SPS slope over a range of angular scales from 16\arcmin to 20$\degree$. Due to GALFA's complex observing and calibration strategy, we construct detailed estimates of the noise contribution and  account for GALFA beam effects on SPS. This allows us to systematically analyze \HI\ images that trace a wide range of interstellar environments.

We find that SPS
slope varies between $\sim −2.6$ at high Galactic latitudes, and $\sim −3.2$ close to Galactic plane. The range of SPS slope values becomes tighter when we consider \HI\ optical depth and line-of-sight length caused by the plane-parallel geometry of \HI\ disk. This relatively uniform, large-scale distribution of SPS slope is suggestive of large-scale turbulent driving being a dominant mechanism for shaping \HI\ structures in the MW and/or the stellar feedback turbulence being efficiently dissipated within dense molecular clouds. Only at latitudes above $60$ degrees we find evidence for \HI\ SPS slope being consistently more shallow. Those directions are largely within the Local Bubble, suggesting the recent history of this cavity, shaped by multiple supernovae explosions, has modified the turbulent state of \HI\ and/or fractions of \HI\ phases.
\end{abstract}

\keywords{\href{http://vocabs.ands.org.au/repository/api/lda/aas/the-unified-astronomy-thesaurus/current/resource.html?uri=http://astrothesaurus.org/uat/833}{Interstellar atomic gas (833)}; 
\href{http://vocabs.ands.org.au/repository/api/lda/aas/the-unified-astronomy-thesaurus/current/resource.html?uri=http://astrothesaurus.org/uat/847}{Interstellar phases (850)};
\href{http://vocabs.ands.org.au/repository/api/lda/aas/the-unified-astronomy-thesaurus/current/resource.html?uri=http://astrothesaurus.org/uat/847}{Interstellar medium (847)}; 
\href{http://vocabs.ands.org.au/repository/api/lda/aas/the-unified-astronomy-thesaurus/current/resource.html?uri=http://astrothesaurus.org/uat/1054}{Milky Way Galaxy (1054)};
\href{http://vocabs.ands.org.au/repository/api/lda/aas/the-unified-astronomy-thesaurus/current/resource.html?uri=http://astrothesaurus.org/uat/1099}{Neutral hydrogen clouds (1099)} 
}

\section{Introduction} \label{sec:intro}
    Interstellar turbulence is an integral driver in structuring the interstellar medium (ISM) \citep[e.g.,][] {2004ARA&A..42..211E, 2007ARA&A..45..565M, 2009SSRv..143..357L}. It is thought to be ubiquitous throughout the ISM and important for many astrophysical processes, such as molecule formation \citep{2017ApJ...843...92B}, accretion within disks around planets \citep{2010A&A...520A..17K}, and even large-scale galactic dynamics \citep[e.g.,][] {2010A&A...520A..17K, 2016MNRAS.458.1671K}. Many astronomical sources and processes can drive interstellar turbulence, which then cascades and affects distributions of interstellar gas and dust.

    Over the past two decades, numerous observational \citep[e.g.,][]{1983A&A...122..282C, 1999MNRAS.302..417S} and numerical studies \citep[e.g.,][]{2017MNRAS.466.1093G, 2018MNRAS.479.3167G} have investigated turbulent properties in the ISM. Despite these efforts, many questions related to the physical processes driving interstellar turbulence remain unanswered. These include the scales on which these processes operate and the observational signatures they imprint on the structure of interstellar clouds over their lifetime. Additionally, the characterization of turbulent properties remains a challenge, with open questions regarding their manifestation across different ISM phases, such as the cold, unstable, and warm neutral medium (CNM, UNM, and WNM).  Similarly, the influence of optical depth ($\tau$) and line-of-sight depth also remain largely unconstrained.

    Many different sources of turbulence have been suggested over the years. They all influence turbulence properties (the slope of fluctuations, driving-scale and modes of dominant motions) in different ways. For example, stellar feedback sources (proto-stellar jets, stellar outflows and winds, photo-heating, supernova explosion) operate largely on sub-cloud spatial scales \citep{2001JKAS...34..333K, 2005A&A...436..585D, 2006ApJ...653.1266J,2009AJ....137.4424T,2011ApJ...731...41O, 2012MNRAS.425..720S, 2013RMxAA..49..137F, 2017MNRAS.466.1093G}, while gravitational and magneto-rotational instabilities \citep{2002ApJ...577..197W, 2010ApJ...718L...1B, 2016MNRAS.458.1671K} operate largely on cloud-size spatial scales. In addition, the supernova-driven turbulence is more effective in producing compressive motions, while in regions with low star formation activity solenoidal motions are expected to be more prominent \citep{2010A&A...512A..81F}.

    Although the various sources of turbulence affect the density and velocity fluctuations in three dimensions, ISM turbulence is traditionally studied using statistical descriptors of intensity fluctuations. For instance, the spatial power spectrum, SPS, \citep{1983A&A...122..282C,2013ApJ...779...36P, 2018ApJ...856..136P}, the $\delta$-variance \citep{1998A&A...336..697S}, and different orders of structure functions \citep{2012ApJ...759L..27P, 2015ApJ...805..118B} are frequently used. The SPS, in particular, is widely applied in both Galactic and extra-galactic studies \citep{2001ApJ...547..792D, 2001ApJ...548..749E, 2010ApJ...718L...1B, 2012A&A...539A..67C, 2012ApJ...751...77Z, 2013ApJ...779...36P, 2019ApJ...887..111S}. The properties of intensity fluctuations can be used to derive information about underlying velocity and/or density fluctuations, as well as to uncover the spatial scales at which energy is injected or dissipated and the mechanisms by which this energy cascades from large to small scales, or vice versa.

    There have been many studies of the \HI\ turbulent properties in the MW using the SPS.
    For example, the SPS method was applied on \HI~images for various regions of our Galaxy \citep{1983A&A...122..282C, 1993MNRAS.262..327G, 2001ApJ...561..264D, 2013ApJ...779...36P, 2018ApJ...856..136P}.
    Similarly, \HI observations of several nearby galaxies, such as the Small Magellanic Cloud \citep[SMC;][]{1999MNRAS.302..417S,2001ApJ...551L..53S,2019ApJ...887..111S}, the Large Magellanic Cloud \citep[LMC;][]{2001ApJ...548..749E,2019ApJ...887..111S}, galaxies from the LITTLE THINGS survey \citep{2012AJ....144..134H,2012ApJ...754...29Z} have been investigated with the SPS. The results of these studies have shown the existence of a hierarchy of structures in the diffuse ISM, with the SPS spectral slope varying slightly with interstellar environments. For example, \HI~integrated intensity images of several regions in the MW, which are largely confined to the disk and likely affected by high optical depth, showed a power-law slope of the SPS of $\sim −3$ \citep{2001ApJ...561..264D}. A slightly steeper slope of $−3.4$ was found for the \HI\ column density in the SMC, and $−3.7$ for the LMC. Galaxies like the LMC and M33 showed a break in their \HI~power spectra, while no departure from a single power law was observed in the MW, the SMC \citep{1999MNRAS.302..417S,2019ApJ...887..111S, 2020MNRAS.492.2663K}, and the Magellanic Bridge when the integrated \HI~intensity distributions were examined. \cite{2004ApJ...616..845M} did notice a change in the SPS slope across several different regions between $−2.9$ and $−3.5$. 
 
  These observational studies offer valuable insights into the distribution of power at different spatial scales, which can be compared with theoretical and numerical predictions to determine the characteristics of the turbulent cascade \citep{2010ApJ...718L...1B,2011MNRAS.417.2891P,2017MNRAS.466.1093G}. The SPS can also identify the significance of shocks \citep{2005ApJ...624L..93B}, and the scales at which energy is injected and dissipated \citep{, 2007ApJ...666L..69K}, therefore providing
crucial constraints for numerical simulations.
    

    In this study we continue our investigation of spatial variations of turbulent properties as a way of quantifying the importance of stellar feedback, gravity, and ram pressure as structuring agents of the \HI~distribution in  the MW. In particular, to investigate possible spatial variations of the SPS slope we apply a consistent methodology on the \HI~data from the entire Galactic Arecibo L-Band Feed Array-\HI (GALFA-\HI) survey \citep{2011ApJS..194...20P}, which offer a unique compromise between a high angular resolution and a large spatial area (resulting in a high spatial dynamic range).

    This paper is organized in the following way. In Section \ref{sec:data} we summarize the data sets used in this study. Section \ref{sec:sps_method} explains our SPS and rolling SPS approach for mapping spatial variability of turbulent properties across the GALFA-\HI~survey. 
    We discuss sources of noise and present a method to characterize the noise contribution to the SPS in the Appendix section. Section \ref{sec:results} presents our SPS slope image and considers how optical depth and plane-parallel geometry affect \HI~turbulent properties. Finally, we discuss and conclude our results in Section \ref{sec:summary}.    
    



\section{Data: GALFA-\HI}\label{sec:data}

    The primary data utilized in this investigation originates from the second data release (DR2) of the GALFA-\HI\ survey \citep{2018ApJS..234....2P}, which was carried out using the ALFA seven-beam receiver array. The full width at half maximum (FWHM) size of the elliptical ALFA beam is $3.3\arcmin \times 3.8 \arcmin$. The survey area spans approximately 13,000 square degrees, 
    including the declination range (DEC) of $-1\degree 20\arcmin$ to $38\degree 02\arcmin$, and the entire range of right accession (RA). The survey 
    covers the velocity range of $-$650 to $+$650 km s$^{-1}$, with velocity resolution of 0.18 km s$^{-1}$.  Further details about the GALFA-\HI\ survey observations and data processing  are provided in \cite{2018ApJS..234....2P}. 
    
    The full velocity coverage
    includs the local MW \HI\, as well as intermediate and high-velocity \HI\ beyond what is expected from the MW rotation curve. To focus solely on 
    the local \HI\, or the  low-velocity clouds (LVC), we limit the velocity range to $|\rm{v}|\leq$ 40 km s$^{-1}$.
    We calculate the \HI\ column density under the assumption of \HI\ being optically thin.
    The \HI\ column density, $N(HI)_{\rm thin}$, is given by:

\begin{equation}\label{eq:HI_thin}
N_{HI,\rm{thin}} = C_0 \, \int T_B(v) dv.
\end{equation}      
    where   $C_0 = 1.823 \times 10^{\rm{18}}$ cm$^{-2}$/(K km s$^{-1}$) \citep{2011piim.book.....D}, $dv$ is the width of a single spectral channel in km s$^{-1}$, and $T_{B}$ in K is the brightness temperature.
    Equation (\ref{eq:HI_thin}) is used to approximate $N(HI)$ in the absence optical depth information. In Section \ref{subsec:optdep} we investigate the effect of high optical depth on the SPS slope.

\begin{figure}[!ht]
    \includegraphics[scale=0.5]{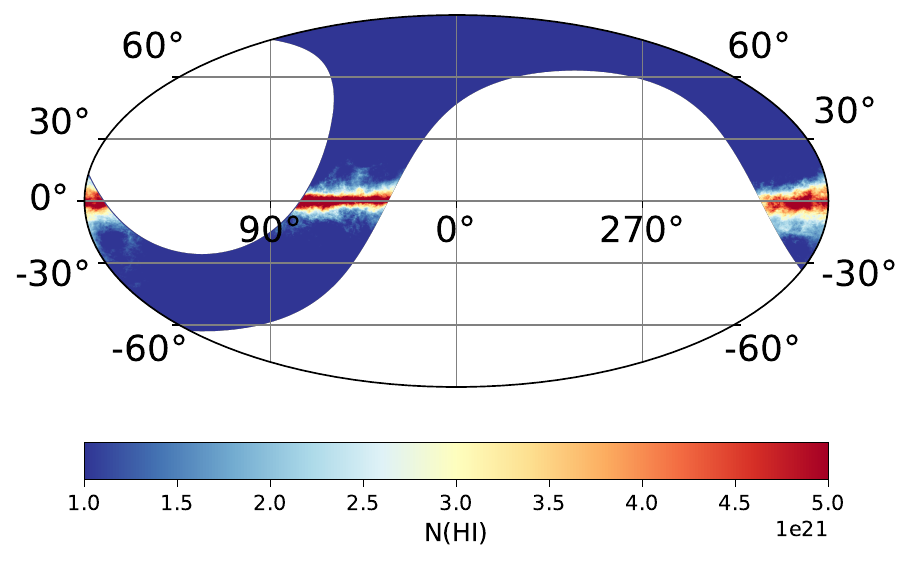}
    \caption{The \HI\ column density calculated by integrating the \HI\ brightness temperature from the GALFA-\HI\ survey over the velocity range $-$40 $\leq \rm v \leq$ 40 km s$^{-1}$ . The Galactic coordinate grid is overlaid.}
    \label{fig:galfa_hpx}
\end{figure}

    Figure \ref{fig:galfa_hpx} shows the spatial coverage of the GALFA-\HI\ survey.
    The survey includes two regions of the MW plane, inner and outer Galaxy, as well as high-latitude \HI. 
     
\section{Methods: Spatial power spectrum (SPS) and the rolling SPS} \label{sec:sps_method}

\subsection{SPS} \label{subsec:sps}

    To calculate \HI\ SPS we use a methodology similar to these from earlier studies \citep{1983A&A...122..282C,1993MNRAS.262..327G,1999MNRAS.302..417S,2000AAS...19711206E}. For a 2D image $I(x)$, the 2D SPS is defined as:    
\begin{equation}\label{eq:powerspectrum}
        P(k) = \iint \langle I(x) I(x') \rangle e^{-i{{\bf L}\cdot{\bf \emph{k}}}}\, dL
\end{equation}
    where $k$ is the spatial frequency, measured in units of wavelength and being proportional to the inverse of the spatial scale\footnote{ Spatial frequency ($k$ in arcmin$^{-1}$) varies inversely with spatial scale ($\lambda$ in arcmin) as: $k = 1/\lambda$.}, while 
    ${\bf L} = {\textbf{\emph{x}}}-{\textbf{\emph{x}'}}$ is the distance between two points.

    We first regrid the GALFA-\HI\ column density image to ensure independent pixels, with the pixel size matching the angular size of the telescope beam which is $\sim4'$. We then apply a median filter with box size of 1$^{\circ}$ to identify pixels with column density $\geq$5-$\sigma$, where standard deviation was calculated for the same 1$^{\circ}$ box. We replace the hot/extreme pixels with the corresponding values from the 
    median filter image.   
    This is done to remove $\sim$0.005$\%$ hot pixels which would affect the Fourier transform due to sharp discontinuities.

    To calculate the SPS for an image (column density), we take the Fourier transform and square the modulus of the transform, $\langle \Re^2 +\Im^2 \rangle$ (where $\Re$ and $\Im$ are the real and imaginary parts of the 2D Fourier transform), which we refer to as the modulus image.
    We proceed by selecting a range of annuli within the modulus image, with evenly spaced intervals determined by $0.03 \log(k)$. This specific bin size strikes a balance, as both increasing and decreasing this spacing have drawbacks: an increase would lead to a reduction in the number of data points in the fitting, while a decrease would result in the appearance of only a few bins near the center of the Fourier plane. Within each annulus, we assume azimuthal symmetry and compute the median value. \cite{2013ApJ...779...36P} demonstrated that the median provides a more accurate representation of the average power due to occasional bright pixels in the modulus image caused by the Gibbs phenomenon. The Gibbs phenomenon, also known as the edge effect, arises due to sharp image boundaries and the \HI\ emission rarely reaching zero at the edges of the surveyed regions. As the Fourier transform of a step function is a $sinc$ function \citep{bracewell1986fourier}, bright pixels appear at the center of $P(k)$  (an example is shown in Figure 2 of \cite{2018ApJ...856..136P}). 
    
    We plot the median power as a function of $\log$ (spatial scale), and estimate the uncertainties using the median absolute deviation (MAD) divided by $\sqrt{N}$, where $N$ is the number of independent pixels in each annulus
    As most previous studies of the \HI\ SPS in the MW have shown featureless SPS spectra \citep{2013ApJ...779...36P,2021ApJ...908..186M}, we start by fitting a single power-law function to the observed SPS:   
\begin{equation}\label{eq:ideal_sps}
 P_{\rm obs}(k) \propto  P_{\rm model} = A \times k^{\gamma}
\end{equation}
    where $A$ is the power-law amplitude, and $\gamma$ is the power-law SPS index for a given SPS.
    When fitting the SPS we use a weighted fitting procedure where weights for each binned data points are calculated as inversely proportional to the square of the uncertainty for that  binned data point \citep{1997ieas.book.....T}.

\subsection{Modeling of the observed SPS}\label{subsec:model_SPS}
    Equation (\ref{eq:ideal_sps}) is an ideal model which does not consider any telescope systematic effects.  
    However, as pointed out by several recent studies \citep{2022PASA...39....5P,2021ApJ...908..186M,2020MNRAS.492.2663K,2019A&A...627A.112K,2017ApJ...834..126B, 2015ApJ...809..153M}, an observed SPS is affected by the telescope beam and fluctuations due to instrumental noise (random and systematic). Therefore, we adopt the methodology proposed by \cite{2020MNRAS.492.2663K} to account for instrumental effects:   
\begin{equation} \label{eq:obs_sps}
    P_{\rm{obs}}(k)  = P_{\rm{beam}}(k) \times P_{\rm{model}}  +  P_{\rm{noise}}(k) 
\end{equation}
    where we assume a 2D Gaussian beam shape with a FWHM $\rm{4}\arcmin \times \rm{4}\arcmin$ size, therefore $P_{\rm{beam}}(k) \propto \exp(-4\pi^2\sigma_{\rm{beam}}^{2}k^2)$, and $\sigma_{\rm{beam}} = \rm{FWHM}/2\sqrt{\rm{\log2}}$.  
    In the case of GALFA-\HI, the noise term $P_{\rm{noise}}(k)$ is not just due to random (white) noise and therefore can not be represented as a constant term in Equation (\ref{eq:obs_sps}). In fact, the noise comes from several contributions, including residuals from the telescope scanning pattern, and we show in the Appendix that the noise term can be written as
    $P_{\rm{noise}}(k) = C\times Q(k,sd)$ as it because it depends on the measured standard deviation of the total noise column density image and brightness temperature.
    To estimate the noise power spectrum $Q(k,sd)$, we first generate a noise image that takes into account both the off-\HI\ line noise (modeled as white Gaussian noise) and systematic noise resulting from scanning artifacts.  From the noise image we calculate the noise power spectrum 
    and then parameterize it as $C\times Q(k,sd)$ to make the noise power spectrum smoother and improve statistics. The Appendix contains a comprehensive description and modeling of  the noise power spectrum. 
    We also show there several examples of the noise SPS and its contribution to the \HI\ SPS for different Galactic regions.

    To fit the SPS in Fourier space, we only consider angular scales where $k\leq k_{\rm {max}}$, which is defined as four times the FWHM. The purpose of this limitation is to avoid spatial scales where the complex ALFA beam pattern (which we model as a 2D Gaussian function) may still affect the SPS.    Conversely, we restrict the fitting process at the larger spatial scale end to  $k>k_{\rm{min}}$, which is defined as the inverse of half the size of the image. This is done because the largest scale bin is estimated from just a few samples in the 2D power spectrum and thus has a large uncertainty.
    
    Prior to fitting the SPS, we compare $P_{obs}(k)$ and $P_{noise}(k)$ and proceed with fitting a power-law only if $P_{obs}(k) > 9\times P_{noise}(k)$ for the entire range of spatial frequencies.
    This is equivalent to ensuring that the signal-to-noise ratio (S/N) exceeds 3 in the image domain. As a result, some pixels in the SPS slope image (Figure \ref{fig:GALFA_LVC_SPS_ERR}) are left blank, those are the pixels that do not satisfy the noise threshold criterion.

\subsection{Rolling SPS} \label{sec:rsps}
    We use the recently developed method -- the rolling SPS -- by \cite{2019ApJ...887..111S} which runs a moving kernel across an \HI\ column density image to produce an image of the SPS slope.  To ensure the best possible angular resolution while having enough statistics to calculate the SPS we selected a 3$\degree$ kernel size.
    In their study, \cite{2019ApJ...887..111S} examined the optimal kernel size by analyzing simulated images. They found that increasing the kernel size led to a decrease in the standard deviation of the mean value of the SPS slope, as demonstrated in Figure A of their appendix. Additionally, they observed that increasing the sub-image size resulted in a reduction of usable pixels at the edge of the image, which conveyed less information about small-scale structure variations.
    
    In this study, we probed angular scales in the range of $1.5 \degree$ to $16\arcmin$ per single 3-degree kernel. To obtain the SPS slope $\gamma$ for each kernel, we follow the methodology outlined in the previous section, accounting for the effects of both the beam and noise. We then shift the kernel by $8\arcmin$ and repeat the process until the SPS slopes are computed across the entire GALFA-\HI\ field.
    During the SPS fitting process we also check the goodness of the fit by evaluating the $\chi^2$ for each fit.


\section{Results}\label{sec:results} 

\subsection{GALFA-\HI\ for the inner, mid and outer Galactic regions} \label{subsec:sps_40}

\begin{figure*}
    \centering
    \subfloat{\includegraphics[scale=1.0]{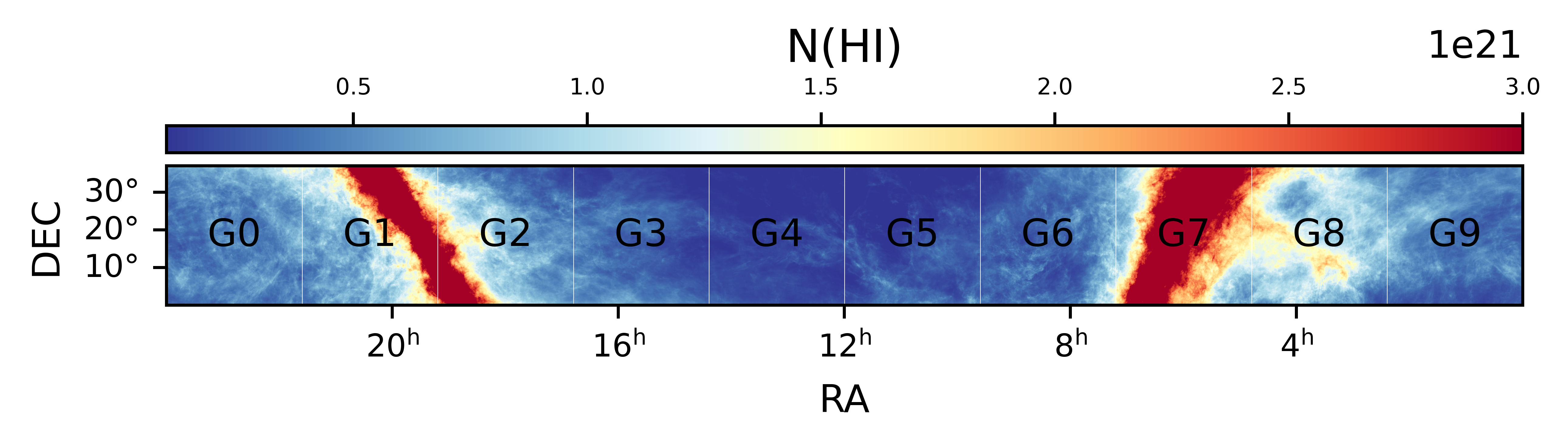}}
    
    \subfloat{\includegraphics[scale=0.75]{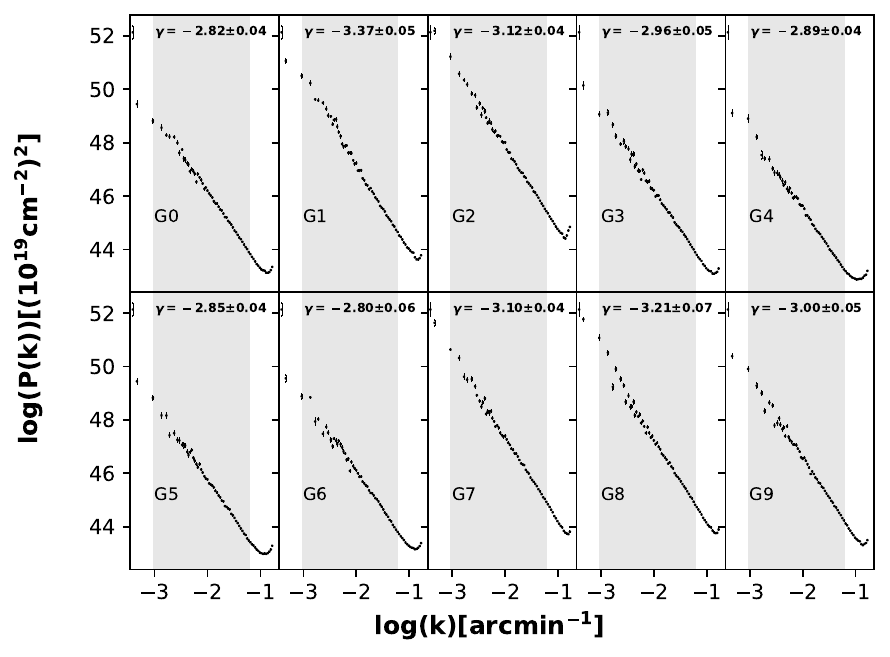}}
    \caption{Top: The GALFA-\HI\ LVC column density image, divided into ten equal sub-regions named as G0 to G9 where each region is $36 \times 36$ degrees in size. 
    Bottom: The corresponding \HI\ SPS calculated for the above G0 to G9 regions. The grey area shows the range of angular scales used for fitting the SPS (between 18$\degree$ and 16$\arcmin$, which corresponds to spatial frequencies $k=10^{-3}$ to $\sim 6\times 10^{-2}$ arcmin$^{-1}$).}
    \label{fig:GALFA_LVC_HI_9}
\end{figure*}

    To investigate spatial variations of \HI\ turbulent properties we start by dividing the entire \HI\ column density image into ten equal regions, each about $36 \times 36$ degree in size.
    These regions are shown in the top panel of Figure \ref{fig:GALFA_LVC_HI_9} and sample the inner MW plane \HI\ emission, the high latitude, 
    and the outer MW plane \HI\ emission, respectively.
    The bottom panels of Figure \ref{fig:GALFA_LVC_HI_9}
    show the SPS of these regions. In fitting the SPS, we used the approach explained in Section \ref{sec:sps_method}. The SPS shown in Figure \ref{fig:GALFA_LVC_HI_9} was fitted  over the angular range of scales 18$\degree$ to $16\arcmin$. 
    
    Figure \ref{fig:GALFA_LVC_HI_9} shows that for all regions the \HI\ SPS is well represented with a single power-law function without obvious break points. It also suggests that the inner and outer disk regions (viewed mainly in regions G1 and G7) have a slightly steeper SPS slope  $\sim -3.2 \pm0.1$ relative to the high-latitude regions (G4 and G5) which have a more shallow slope, $\sim -2.9 \pm0.1$. 
    In particular, the two most extreme slopes are found for regions G1 and G6, with corresponding SPS slopes being 
    $-3.37 \pm0.05$ and $-2.80 \pm0.06$ respectively.
    This result suggests that 
    some spatial variations of the SPS slope may exist across the MW.

\begin{figure*}
    \centering
    \subfloat{\includegraphics[scale=1.0]{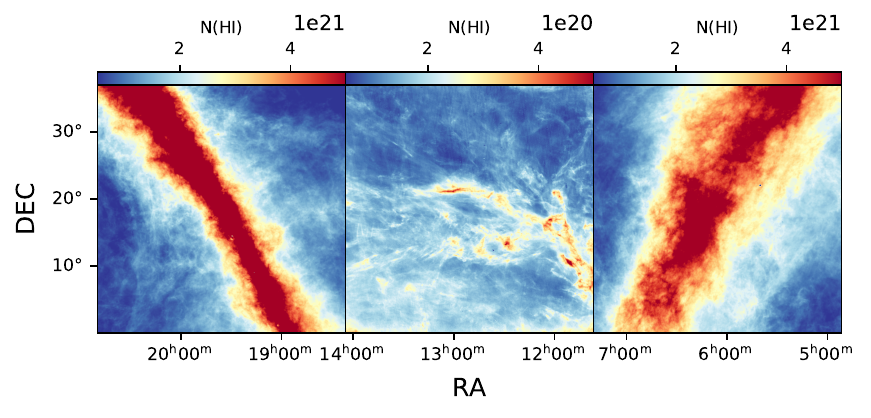}}
    \caption{The \HI\ column density images for three regions: inner Galactic plane (left), high Galactic latitude (mid), and outer Galactic plane (right). Each region has a size $40 \times 40$ degrees. The three regions are centered at:
    RA $\sim$ 19h30m (left), RA $\sim$ 12h45m (mid), and RA $\sim$ 06h00m (right) with declination centered at $18.5$ degree for each.}
    \label{fig:GALFA_LVCHI_40}
\end{figure*}

    To demonstrate the difference in the SPS slope between the disk and high-latitude \HI\ and to extend the range of angular scales even further, in Figure \ref{fig:GALFA_LVCHI_40} we select three regions, each about $40 \times 40$ degrees large, specifically centered on the inner Galaxy, the outer Galaxy, and at high latitudes. 
 The inner Galaxy region covers the longitude range 30 to 77 degrees and for our selected velocity range includes mostly the Perseus arm (and a small fraction of \HI\ in each of the near Sagittarius-Carina and the Outer Scutum Centarus arms),
therefore including the \HI\ mostly within few to 10 kpc from the Sun. The outer Galaxy region covers the longitude range of 170 to 215 degrees, and includes four major spiral arms 
(Local, Perseus, Scutum-Centaurus and the Norma–Outer arm which is at $\sim10$ kpc from the Sun). 
While this region includes the \HI\ 
within up to 5-10 kpc from the Sun, it also includes some more distant \HI\ that reaches Galactocentric distance of up to 20 kpc, as opposed to the inner Galaxy region which stays within the Galactocentric radius of 10 kpc.
    
\begin{figure}%
    \centering
   
    \subfloat{\includegraphics[scale=0.5]{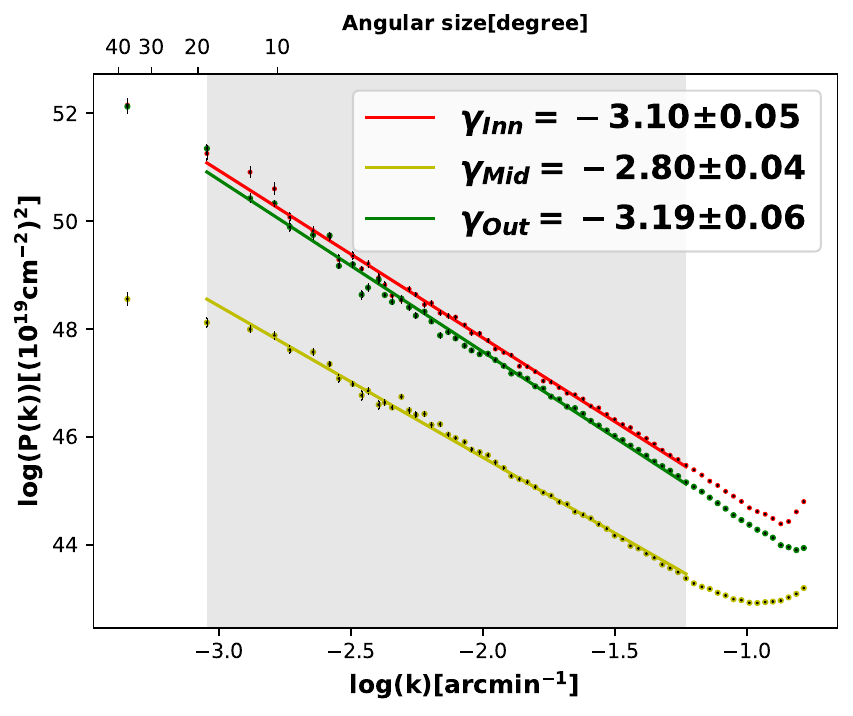}}
    \caption{The SPS for three \HI column density images shown in Figure \ref{fig:GALFA_LVCHI_40}. The fitting range, shown by the grey area, covers a  range of angular scales from $\sim20$\degree ~to 16$\arcmin$,
    which corresponds to spatial frequencies $k=10^{-3}$ to $\sim 6\times 10^{-2}$ arcmin$^{-1}$.} 
    \label{fig:GALFA_LVC_40_SPS}
\end{figure}

    The corresponding SPS for the three regions is shown in Figure \ref{fig:GALFA_LVC_40_SPS}. This figure confirms that the inner and outer Galactic plane regions (shown as red and green lines) have similar SPS slopes that are consistently slightly steeper than the slope of the high-latitude region (shown in yellow). 
     If we divide the fitting range into two, the three regions have consistent SPS slope measured for the small-scale end of the fitting range. However, on the large-scale end of the fitting range, the high-latitude region has a much shallower SPS slope relative to the low-latitude regions (e.g. $-2.35 \pm 0.17$ vs $-3.73 \pm 0.17$). This shows that
    the main difference between low- and high-latitude \HI\ SPS occurs on scales larger than $\sim5$ degrees. This is due to lack of large-scale diffuse \HI\ emission at high latitudes. The same conclusion can be reached by comparing top two panels of Figure \ref{fig:4POS_SPS}in the Appendix, where the noise SPS between low- and high-latitude regions is comparable, but the \HI\ power on large scales is a factor of $\sim100$ higher for the low-latitude field.  It is also interesting to find that the inner and outer Galaxy regions have consistent SPS slopes. While both regions probe \HI\ within a few kpc from the Sun, the outer Galaxy region is at a higher Galactocentric radius.

\subsection{Rolling SPS for the entire GALFA-\HI\ field}

   The rolling SPS analysis, as explained in Section \ref{sec:sps_method}, is employed using a 3-degree kernel over the entire low-velocity \HI\ column density image. The SPS slope is calculated for each kernel, but only for regions where the \HI\ brightness temperature is at least three times higher than the noise level to avoid areas dominated by noise.  As we explained in Section \ref{subsec:model_SPS}, the S/N check is done in the Fourier plane, and we do not perform fitting if the S/N threshold is not met.

\begin{figure*}
    \centering
    \subfloat{\includegraphics[scale=1.2]{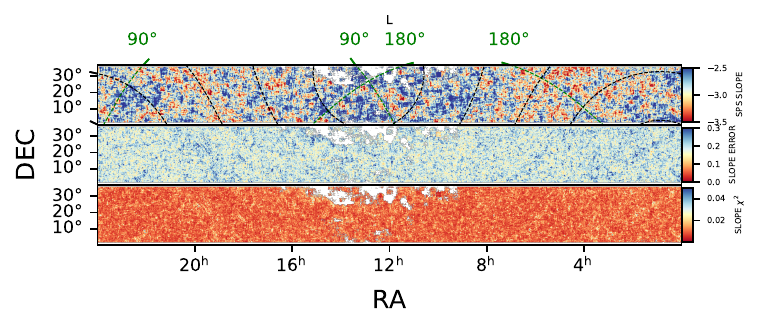}}
    
    \caption{Top: The GALFA-HI rolling SPS slope, where the SPS slope is calculated for each $3\degree$ kernel and rolled across RA and DEC by stepping every $8\arcmin$ to cover the entire GALFA-HI region. The SPS slope was determined only for regions where the SPS power $\geq 9 \times$ noise power, and the fitting was conducted over the range of 1.5$\degree$ to 16$\arcmin$. Pixels that do not satisfy this criterion are left blank (NaN). The Galactic coordinate grid is overlaid as black dashed lines.  Middle: The uncertainty in the fitted SPS slope. Bottom: The $\chi^2$ map of the single power-law SPS fit.}
    \label{fig:GALFA_LVC_SPS_ERR}
\end{figure*}

   Figure \ref{fig:GALFA_LVC_SPS_ERR} (top) displays the resulting rolling SPS slope image, with noise-dominated pixels being left blank
  (that way these pixels are not included in further analyses).
The uncertainty in the estimated slope is shown in the middle panel of the figure, with a typical uncertainty of 0.1-0.2. No residual structure is visible in this image. Finally the bottom panel of the figure shows the $\chi^2$ image which looks relatively uniform and featureless and  we conclude that a single power-law fit provides a good fit to the SPS and discard any need of broken power-law fitting.

\begin{figure}[h]
    \centering
    \subfloat{\includegraphics[scale=0.5]{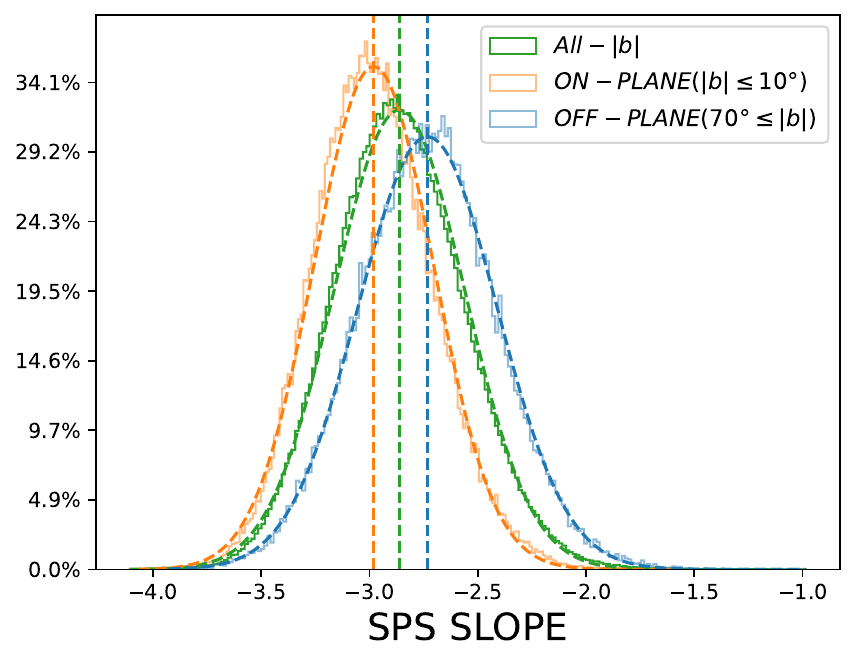}}
    \caption{The histogram (shown in green) of the SPS slope image shown in the top of Figure \ref{fig:GALFA_LVC_SPS_ERR}.
    The on-plane histogram corresponds to latitudes $|b|\leq 10 \degree$ is shown in blue, whereas the off-plane histogram pertains to latitudes $|b|\geq 70 \degree$ and is shown in orange.The vertical dashed lines represent the mean value for the three samples.}  
    \label{fig:GALFA_SPS_hist_LVC}
\end{figure}

    On a 3-degree spatial scale, the SPS slope varies between $-2.6$ and $-3.2$ (1-$\sigma$ range). This figure confirms that the Galactic plane regions have systematically the steepest SPS slope (the image is dominated by red/orange color), while the high Galactic latitude regions have the most shallow SPS slope (the image is dominated by blue color). In Figure \ref{fig:GALFA_SPS_hist_LVC} we plot histograms of pixels shown in the top of Figure \ref{fig:GALFA_LVC_SPS_ERR} for the most extreme SPS slope values: in orange we show data for $|b|\leq10$ degrees, while in blue we show data for $|b|\geq70$ degrees. The mean SPS slope for these two sub-samples is $\sim$ $-3.0$ and $\sim$ $-2.7$, respectively. As we have already seen, there appears to be a slight difference in SPS slope values between the MW plane and high latitude regions. However, Figure \ref{fig:GALFA_LVC_SPS_ERR} also shows that pixels with low signal/noise (S/N) are mainly found at high Galactic latitude. We excluded extreme, low-S/N pixels while fitting the SPS. As we discussed in the Appendix, the GALFA-\HI\ noise is complex and affects a range of angular scales. If not accounted properly (see Figure \ref{fig:4POS_SPS}), it would result in a more shallow SPS slope. While some of the trend in the SPS slope being more shallow at high latitudes could be due to the systematic decrease of S/N as we move away from the Galactic plane, we have experimented with changing the range of scales used for fitting. We find that the shallow SPS slope at high latitudes persists and can not be explained as being caused by the noise SPS in low S/N regions.

    In Figure \ref{fig:slope-comparison}
    we show the SPS slope values as a function of Galactic latitude. For each 2 degree wide Galactic latitude bin the average SPS slope value is calculated from the SPS map shown in Figure \ref{fig:GALFA_LVC_SPS_ERR} (top panel). Since the rolling SPS method uses a 3-degree wide kernel, shifted by 8 arcmins, the SPS slope values in Figure \ref{fig:GALFA_LVC_SPS_ERR} are correlated. The error bars in Figure \ref{fig:slope-comparison} are estimated from the standard deviation of SPS slope values within each 2 degree Galactic latitude range and divided by $\sqrt{(N'-1)}$. Here, $N'$ is an estimate of the number of independent kernels within each 2-degree Galactic latitude zone, given by the number of pixels in Figure \ref{fig:GALFA_LVC_SPS_ERR} that fall within each 2-degree latitude bin divided by the number of pixels per kernel, which is (180/8)$^{\rm 2}$.

    Figure \ref{fig:slope-comparison} shows that the SPS slope remains roughly constant at $\sim-2.9$ across the entire latitude range. For $|b|>60$ degrees, the slope becomes slightly more shallow. Our mean SPS slope is in excellent agreement with what \cite{2019A&A...627A.112K} found by using the all-sky HI4PI survey. Using the entire data set 
    they found a SPS slope of $-2.943 \pm 0.003$ and $-2.944 \pm 0.005$ when considering only $|b|>20$ degrees (for the velocity range from $-8$ to 8 km s$^{-1}$). 
    Figure \ref{fig:slope-comparison}    
    hints at a slight asymmetry between the northern and southern hemispheres. While on the north side we see a gradual change in the SPS slope from $b\sim0^{\circ}$ to $b\sim80^{\circ}$, on the southern side the SPS slope is relatively constant from $b\sim0^{\circ}$ to $-50^{\circ}$, and then has a sharp decrease from $\sim-2.9$ to $\sim-2.6$. This asymmetry in SPS slope between north and south latitudes likely stems from the asymmetric \HI\ distribution in the MW \citep{1988gera.book..295B,2006ApJ...643..881L, 2006Sci...312.1773L, 2007A&A...469..511K, 2009ARA&A..47...27K}.

 \subsection{Comparison of the SPS slope with previous studies}

\begin{figure*}
    \centering
    \subfloat{\includegraphics[scale=0.75]{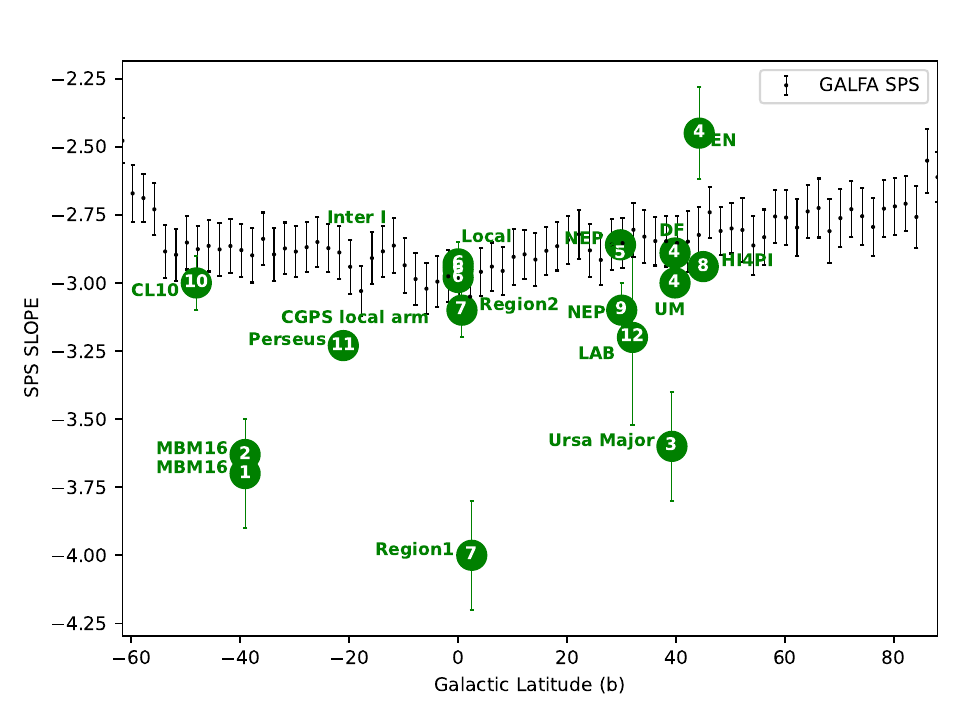}}
    \caption{The \HI\ SPS slope, averaged in 2$^{\circ}$ latitude bins, as a function of Galactic latitude. Green points show results from previous studies: 
    1- \cite{2013ApJ...779...36P},
    2- \cite{2017A&A...607A..15K},
    3- \cite{2003A&A...411..109M},  
    4- \cite{2017ApJ...834..126B},
    5- \cite{2015ApJ...809..153M}, 
    6- \cite{2006ApJS..165..512K}, 
    7- \cite{2001ApJ...561..264D}, 
    8- \cite{2019A&A...627A.112K},
    9- \cite{2021ApJ...908..186M},
    10- \cite{2010ApJ...714.1398C},
    11- \cite{2018ApJ...856..136P}, 
    12- \cite{2019MNRAS.483.3437C}.}
     \label{fig:slope-comparison}
\end{figure*}

    In Figure \ref{fig:slope-comparison} we also compare our SPS slope with results from previous studies. There have been many studies of the \HI\ SPS mostly focusing on smaller regions. While most studies shown in this figure have used the Fourier transform method to calculate the SPS, the exact details of the methods used and the fitting range, as well as the velocity range, vary greatly, making a comparison of this heterogeneous sample difficult. Our work is the first to apply the same methodology uniformly over a large spatial area, about 13,000 square degrees.

    We note that all these measurements were obtained from \HI\ ~images integrated over a range of velocity channels that roughly corresponds to the local \HI\ (excluding intermediate and high-velocity gas) and likely represent estimates of density fluctuations \citep{2000ApJ...537..720L, 2021ApJ...910..161Y}.
    Our SPS slope values are generally consistent with many prior findings. Two high-latitude \HI\ clouds, MBM16 and Ursa Major, stand out as having a much steeper SPS slope of $-3.7$, as well as one region in the Galactic plane. MBM16 is especially interesting as it is a starless cloud. Both \cite{2013ApJ...779...36P} and
    \cite{2003A&A...411..109M}
    hypothesized that the steep SPS slope could be caused by the lack of stellar feedback.
    However, it is likely that the smallest spatial scales in these studies were affected by the beam shape and resulted in a steep SPS.
    \cite{2017A&A...607A..15K} re-examined the MBM16 power spectrum and estimated the SPS slope of $-3.63 \pm 0.04$ after applying apodization, beam correction, and the noise bias correction (point 2 in Figure \ref{fig:slope-comparison}). Therefore, the MBM16 region remains as being an example of an unusually steep \HI\ SPS slope. Further studies are needed to understand the origin of this steep slope.
    On the other hand, for the Ursa Major field, 
    \cite{2017ApJ...834..126B} obtained SPS slope of $-2.68\pm0.14$ after accounting for instrumental effects (shown as ``UM'' in Figure \ref{fig:slope-comparison}).

\subsection{Accounting for high optical depth}\label{subsec:optdep}

    When calculating the SPS slope we use the column density images by assuming that \HI\ is optically thin. This assumption is reasonable for high-latitude \HI, but certainly underestimates the \HI\ column density close to and within the MW plane. To investigate whether the SPS slope distribution is affected by the spatial variations of the \HI ~optical depth we applied the correction for high optical depth on the entire GALFA-\HI\ field. As shown by \cite{2015ApJ...809...56L} and \cite{2019ApJ...880..141N}, there are significant regional variations of this correction factor. For example,  \cite{2019ApJ...880..141N} investigated \HI ~in the vicinity of several Giant Molecular Clouds (GMCs) and found that the correction factor ($f$) from their best fit is:   
\begin{equation}\label{eq:opt_depth}
\begin{split}
f &= (0.47 \pm 0.09) \times  \log_{10}(N(HI)/10^{20})\\
& + ( 0.66 \pm 0.12) .
\end{split}
\end{equation}
    However, for sightlines through the Galactic plane area ($|b|<5^{\circ}$), the correction factor increased rapidly as $f= (2.41 \pm 0.93) \times \log_{10}(N(HI)/10^{20}) + (2.57 \pm 1.57)$. 
    At low column densities below $\sim 10^{21}$ cm$^{−2}$, the correction factor was close to unity. 
    
    As a first-order test of how the optical depth correction affects spatial variations of the SPS slope, we applied the correction from Equation (\ref{eq:opt_depth}) on the entire GALFA-\HI\ column density image. This affected the SPS slope only for $|b|<20^{\circ}$  by making the SPS slope slightly more shallow. At latitudes $|b|\geq20^{\circ}$, the SPS slope remained unchanged. The observed change of the SPS slope is, however, within the SPS slope uncertainties (typical error bars shown in Figure \ref{fig:slope-comparison}  are $\sim0.1$). 
    
    We conclude that the optical depth is unlikely to modify significantly the observed relatively uniform large-scale distribution of the SPS slope. While our attempted correction for high optical depth is not realistic at very low latitudes ($|b|<10^{\circ}$), we hypothesize that the full correction would introduce more small-scale structure resulting in a more shallow SPS slope within $10^{\circ}$ from the plane where currently we see the steepest SPS slope (see Figure \ref{fig:slope-comparison}). In essence, the optical depth correction would likely further diminish spatial variations of the SPS slope seen between low and intermediate Galactic latitudes.

\subsection{Accounting for the plane parallel distribution of \HI}

\begin{figure*}
    \centering
    \subfloat{\includegraphics[scale=0.8]{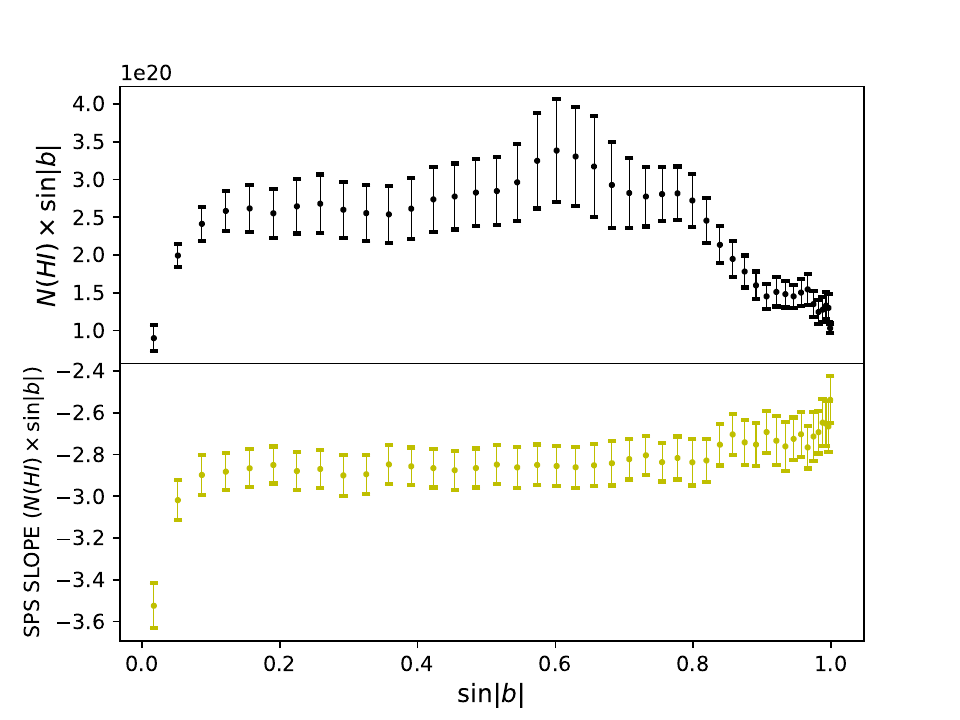}}
    \caption{
    Top panel:  The \HI\ column density integrated over $\pm 40$ km s$^{-1}$ velocity range multiplied by $\sin|b|$  is shown as black data points with respect to $\sin|b|$. 
    Bottom panel: the SPS slope of the $N(HI) \times \sin|b|$ distribution, averaged in 2-degree latitude bins, as a function of $\sin|b|$.}
    \label{fig:GALFA_NHI_SPS_B_SINB}
\end{figure*}

    The \HI\ distribution in the MW has a rough plane-parallel geometry \citep{1990ARA&A..28..215D}. This results in the fact that lines of sight at different Galactic latitudes probe \HI\ along different path lengths. Therefore, the \HI\ column density changes significantly as a function of Galactic latitude.
    We investigate here how the plane-parallel geometry of the \HI\ disk affects the SPS slope.  We therefore multiply $N(HI)$ with $\sin|b|$ as this effectively projects the \HI\ column density along the vertical axis
     and therefore normalizes $N(HI)$ for the difference in path lengths.  
    The $N(HI) \times \sin|b|$ distribution is shown in the top panel of Figure \ref{fig:GALFA_NHI_SPS_B_SINB} as a function of $\sin|b|$, while the rolling SPS slope calculated for the $N(HI) \times \sin|b|$ distribution is shown in the bottom panel of the same figure.  

    The $N(HI) \times \sin|b|$ distribution
    (top panel of Figure \ref{fig:GALFA_NHI_SPS_B_SINB})
    is in agreement with several previous studies and shows several features \citep{1990ARA&A..28..215D,1975AJ.....80..111K}. 
     \cite{1975AJ.....80..111K} used the velocity range of $\pm50$ km s$^{-1}$ and showed that the $N(HI) \times \sin|b|$ distribution is flat at $2.6\times 10^{20}$ cm$^{-2}$ up to $\sin|b|\sim 0.6$, and then decreases roughly by a factor of two (reaching $1.5\times 10^{20}$ cm$^{-2}$ at $|b|=85^{\circ}$).  \cite{1990ARA&A..28..215D} combined data from several \HI\ surveys over the velocity range $\pm250$ km s$^{-1}$ and found that the decrease of $N(HI) \times \sin|b|$ at high latitudes follows: $N(HI)=3.84 \times \csc|b| - 2.11$ in units of $10^{20}$ cm$^{-2}$. 
The excellent agreement between our study and both 
Knapp (1975) and Dickey \& Lockman (1990)
the $N(HI) \times \sin|b|$ distribution 
suggests that our velocity range of $\pm40$ km s$^{-1}$ encompasses the bulk of \HI\ emission and that the decline of the column density at high latitudes is not due to the selected velocity range.   
 
    The $N(HI) \times \sin|b|$ distribution goes to 0 at $b\sim0^{\circ}$  as the MW \HI\ disk has a finite size.
    For $\sin|b|\sim 0.1$-0.8 the $N(HI) \times \sin|b|$ distribution is relatively uniform.
    For $|b|>50^{\circ}$  the \HI\ column density decreases by a factor of $\sim2$ as it is affected by 
    the presence of the Local Bubble which contains largely hot gas 
    \citep{1987ARA&A..25..303C} and some cooler \HI\ \citep{2007arXiv0709.4480R} but this \HI\ distribution is modified by the explosion of multiple supernovae that formed the Local Bubble.  
     While lines of sight at $|b|\leq50^{\circ}$  predominantly probe the disk \HI, at least half of the path length at $|b|>50^{\circ}$  is contained inside the Local Bubble 
    (the scale height of the cold/warm \HI\ is about 300/500 pc near the Sun \citep{2023ARA&A..61...19M}, while the radius of the Local Bubble is $\sim200$ pc \citep{2010A&A...510A..54W})
    and therefore has a different fraction of warm to cold \HI\ relative to the \HI\ at $|b|<50^{\circ}$.

    As shown in the bottom panel of Figure \ref{fig:GALFA_NHI_SPS_B_SINB}, the SPS slope of the $N(HI) \times \sin|b|$ distribution is essentially constant at $ -2.85 \pm 0.02$. This is in agreement with the mostly constant $N(HI) \times \sin|b|$ distribution caused by the plane-parallel geometry of the MW \HI. Only at $\sin|b| >0.8$ where the $N(HI)$ is confined to the Local Bubble and the fraction of cold vs. warm \HI is different,  we do find a small difference, $ -2.72 \pm 0.03$. While this difference is barely at a 3-sigma level, it could suggest the different fraction of \HI\ phases within the Local Bubble and/or different turbulent state of the \HI\ inside the Local Bubble.



\section{Discussion and summary}
\label{sec:summary}

    In this study, we calculated the \HI\ SPS of the \HI\ column density image (integrated over local velocities $\pm 40$ km s$^{-1}$) that covers close to 13,000 square degrees. As the GALFA-\HI\ survey has complex scanning patterns and noise statistics due to its receiver cluster, we developed a methodology to quantify and parameterise the noise (and beam) contributions to the \HI\ SPS across the entire survey area. Using the rolling SPS method, we mapped the SPS slope spatially across the GALFA-\HI\ field by rolling a 3-degree kernel and consistently treating beam and noise contributions to the SPS within each kernel.
 
    We focused only on the \HI\ column density, relying on the assumption that the $\pm 40$ km s$^{-1}$ velocity range depicts the bulk of the \HI\ mass and that the \HI column density is a reliable tracer of 3D density fluctuations. While previous studies \citep[e.g.,][]{2000ApJ...537..720L} have calculated the \HI\ SPS from column density, the question regarding the exact contributions of density vs. velocity fluctuations to the observed intensity fluctuations has received significant attention in recent years. For example, the approach based on varying the width of velocity channels  to separate density from velocity fluctuations \citep{2000ApJ...537..720L} was recently updated with a more complex treatment \citep{2021ApJ...910..161Y}. On the other hand, \cite{2019ApJ...874..171C}  and \cite{2016ApJ...821..117K} suggested that the \HI\ intensity fluctuations seen at different velocities are phase dependent, with narrow velocity channels being dominated by the CNM and causing the \HI\ SPS slope to be more shallow than what is found for the integrated intensity images. 
    In addition, \cite{2021MNRAS.505.1972K} used simulations to show that the \HI\ SPS depends on the multi-phase medium. While understanding of direct contributions from density, velocity, and different \HI\ phases to the \HI\ SPS awaits further development and tests, we emphasize that our approach  provides a systematic treatment of the \HI\ SPS over a range of different environments with varying S/N ratios and therefore enables a comprehensive examination of spatial variations of the \HI\ SPS slope.

    The key result from our study is that we find a relatively uniform SPS slope across most of the field covered by the GALFA-\HI\ survey.  When projecting the \HI\ column density along the vertical axis, by multiplying the \HI\ column density by $\sin |b|$, the SPS slope of the $N(HI) \times \sin |b|$ distribution is remarkably flat (within 1-$\sigma$ error bars) for $|b|\leq60^{\circ}$. Only at high Galactic latitudes, $|b|>60^{\circ}$, we find a slightly more shallow SPS slope. The difference between low and high Galactic latitudes is less pronounced when we consider the high-optical-depth correction and the difference in the line-of-sight length caused by the plane parallel geometry of the \HI\ disk. However, the more shallow SPS slope at $|b|>60^{\circ}$, relative to lower latitudes, remains statistically significant at about 3-$\sigma$ level. 
     At high latitudes the lines-of-sight are mostly within the local ISM where the Local Bubble represents a prominent structure.
 On the other hand, at lower latitudes the lines-of-sight probe much longer path-lengths throughout the Galactic disk. Even after scaling the path-lengths for a direct comparison of the \HI\ column density, it appears that the local ISM has distinct turbulent properties.
     Our  results therefore suggest that \HI\ turbulent properties within the Local Bubble are modified by the multiple supernovae explosions.

    The relatively uniform SPS slope across the majority of the GALFA-HI field is a somewhat surprising finding when considering that the GALFA-\HI\ survey probes many different interstellar environments. For example, the \HI\ column density varies from $\sim10^{20}$ to  $\sim10^{22}$ cm$^{-2}$, tracing regions with diffuse \HI\ all the way to regions with a significant fraction of molecular gas (e.g. near molecular clouds such as Perseus, Taurus and California). Based on \HI\ absorption measurements, it has been shown that the CNM fraction varies across the GALFA-\HI\ field between $\sim0.2$ and 0.75 \citep{2019ApJ...880..141N}. Yet, the SPS slope remains largely around $-2.9$. This may suggest that the phase distribution, fractions of cold vs. warm \HI, have a secondary role in explaining SPS properties. 

    The relatively uniform SPS slope could be caused by the dominance of large-scale turbulent driving, on Galaxy-wide spatial scales, introduced by a combination of Galactic rotation, gravitational instabilities, and/or infall onto the Galaxy \citep{2002ApJ...577..197W,2010ApJ...718L...1B, 2016MNRAS.458.1671K}. Numerical simulations by \citet{2014ApJ...780...99Y} showed that the turbulent energy spectrum is very sensitive to large-scale driving. Unless the energy injection rates on small scales are much higher than the energy injection rate on large scales, the large-scale driving will always dominate. Therefore, even in the presence of significant stellar feedback in the plane of the MW, the large-scale turbulent driving could be the dominant mechanism. The finding of a relatively uniform SPS slope for the MW resembles the \HI\ turbulent properties within  the Small Magellanic Cloud where recently \cite{2019ApJ...887..111S} found a very uniform SPS slope across the entire galaxy. 
    In addition, \cite{2022arXiv220413760Y} applied the new method for separating density and velocity fluctuations from spectral-line observations on \HI\ and CO observations of several Galactic regions and showed that the turbulent velocity cascade has a universal spectrum over a broad range of scales (from $\sim1$ pc to $\sim10^3$ pc). They concluded that this supports the picture where turbulence is driven on large spatial scales, while the energy injection on small spatial scales may not influence significantly the turbulent cascade. This study found that the velocity power spectrum slope was more uniform than the slope of the density fields.

    Furthermore, \citet{2012A&ARv..20...55H} found that the kinetic energy transfer rate, observed in the molecular cloud population traced by $^{12}$CO, exhibited no change in magnitude across a spatial range of $\sim0.01$ to $\sim100$ pc. This led them to conclude that molecular clouds are part of the same turbulent cascade as \HI. If molecular clouds of diverse star formation and stellar feedback levels exhibit negligible changes in their internal turbulent characteristics, it is not surprising that the \HI\ also experiences limited effects from small-scale turbulent driving, which could explain the relatively consistent \HI\ SPS slope. A similar conclusion was reached in \cite{2022ApJ...928..143E} who investigated the \HI\ turbulent properties in a sample of nearby galaxies and  suggested that feedback does not affect much atomic gas, hypothesizing that most of the feedback energy goes into dense molecular clouds where it gets radiated efficiently.

    As shown in Figure \ref{fig:GALFA_LVC_40_SPS}, we do not see any difference between inner and outer Galaxy \HI\ turbulent properties when considering the SPS slope. While the outer Galaxy region includes more distant \HI\, the inner Galaxy region is probing \HI\ largely within a few kpc from the Sun. This is different from \cite{2021A&A...655A.101D} who studied \HI\ turbulent properties of 33 disk galaxies using the $\delta$-variance approach and concluded that inner parts are affected largely by supernova feedback while outer parts are mostly affected by large-scale dynamics. We note that this study has a much lower spatial resolution and has calculated turbulent properties by averaging data for entire galaxies.


\acknowledgments
    This study is supported by the National Aeronautics and Space Administration under Grant No. 4200766703, proposal \cite{2020adap.prop..159S}. We also acknowledge support provided by the University of Wisconsin-Madison Office of the Vice Chancellor for Research and Graduate Education
    with funding from the Wisconsin Alumni Research Foundation. We are very grateful to Daniel Rybarczyk and Trey Wenger for useful discussions at various stages of this project. We are grateful to Peter Kalberla for providing test data and a code for comparison of results. We thank the referee for constructive and insightful comments and suggestions. 


\software{
    Astropy \citep{2013A&A...558A..33A}, 
    DS9 \citep{2003ASPC..295..489J}, 
    HealPy \citep{2005ApJ...622..759G}, 
    KARMA \citep{1996ASPC..101...80G}, 
    Matplotlib \citep{2007CSE.....9...90H}, 
    NumPy \citep{2011CSE....13b..22V}, 
    SciPy \citep{2007CSE.....9c..10O}}

\vspace{1cm}

\newpage

\bibliography{main}{}

\bibliographystyle{aasjournal}


\appendix
\label{appendix}

\section{Noise and telescope beam contributions to the SPS}

\label{sec:noise_Brian}

\subsection{Noise analysis}\label{subsec:noisesigma}

   The calculation of the \HI\ SPS for a given kernel must take into account both the intricate ALFA beam and the noise in the GALFA-\HI\ data, which varies in both the spectral and spatial dimensions. A detailed understanding of the noise is particularly crucial in regions at mid and high latitudes, which represent roughly 25$\%$ of the observed sky, as the signal to noise ratio decreases substantially with Galactic latitude and the noise contribution can impact the slope of the SPS.
   
    The GALFA-\HI\ survey resulted from merging various \HI\ observations obtained via different scanning methods. Although global calibration solutions were derived to address the relative calibration of different regions \citep{2011ApJS..194...20P}, the residuals from distinct scanning patterns persist and constitute a form of systematic noise.
    To account for the noise contribution during the fitting of the rolling 
    \HI\ SPS, we generate a noise image, $N_{\rm{total}}(x,y)$, which serves as the basis for computing the rolling noise SPS, $P_{\rm{noise}}(k)$. The GALFA-\HI\ column density image encompasses multiple noise components, including receiver (rx) noise, noise stemming from the brightness temperature of the incident \HI\ emission (both associated with the system temperature, $T_{\rm{sys}}$), as well as systematic noise originating from calibration artifacts resulting from various scanning techniques employed in the GALFA-\HI\ survey. The calibration noise consists of both an root mean square (rms) component (stemming from the scanning and data merging process) and a systematic offset component (arising from the zero-point calibration). As it is impossible to completely separate the random noise due to receiver temperature ($T_{\rm {rx}}$) from the systematic noise associated with calibration, we choose to divide the total noise image corresponding to the \HI\ column density image into two components:  
\begin{equation}\label{eq:noise3}
    N_{\rm{total}} (x,y) = N_{\rm {off}}(x,y) + N_{\rm {on-HI}}(x,y)
\end{equation}
    where $N_{\rm{off}}$ is the combined contribution of random fluctuations due to receiver temperature plus the systematic scanning calibration uncertainty (this image can be estimated from the emission-free channels of the GALFA-\HI\ data cube), and $N_{\rm{on-HI}}$ is the noise contribution due to the \HI\ brightness temperature.  

{\bf $N_{\rm{off}}(x,y):$}
    The noise from the ALFA receiver, characterized by $T_{\rm{rx}}$ = 30 K \citep{2018ApJS..234....2P}, is independent of velocity and gives rise to the radiometer noise, with a standard deviation of $\sigma_{\rm{rx}}$. To obtain $\sigma_{\rm{rx}}$, we select the velocity range nearly free of \HI\ emission, i.e., $-520$ to $-440$ km s$^{-1}$ (we refer to this subset as the ``non-\HI'' data cube, which shares the same velocity width as our LVC data cube). For each spatial pixel in this cube, we calculate the standard deviation over the velocity axis. This yields the $\sigma_{\rm{rx}}(x,y)$ image (shown in top 3 panels of  Figure \ref{fig:off_HI} ), whose pixel values range from 1.4 to 2.3 K.
    
   We obtain the \HI\ column density of the emission-free data cube by integrating the brightness temperature across the \HI-free velocity range of $-520$ to $-440$ km s$^{-1}$, resulting in the $N_{\rm{off}}(x,y)$ image. The fourth panel of Figure \ref{fig:off_HI} depicts this image, which has pixel values spanning from $-$7.6 $\times 10^{18}$ to $+$7.1 $\times 10^{18}$ cm$^{-2}$ across 97$\%$ of the area. Although it contains receiver noise, its spatial structure is largely determined by the scanning and calibration artifacts. 
    
   Given the challenge of differentiating the systematic calibration noise due to scanning and merging of different observations, we note that the non-\HI\ noise image involves contributions from both the $\sigma_{\rm{rx}}(x,y)$ image (receiver noise) and systematic calibration offsets.

{\bf $N_{\rm{on-HI}}$:}
    The noise term due to the \HI\ brightness temperature is radial velocity dependent and can be expressed by scaling the standard deviation estimated over emission-free channels ($\sigma_{\rm{rx}}$):   
\begin{equation}\label{eq:noise1}
 \sigma_{\rm{rx+HI}} (x,y,v) = \sigma_{\rm{rx}} (x,y) \left (1+T_b(x,y,v)/T_{\rm{rx}} \right )
\end{equation}
    where $T_b(x,y,v)$ is the brightness temperature at each pixel location and each velocity channel. $\sigma_{\rm{rx+HI}}$ is essentially a noise envelope (or spectrum) calculated at each spatial pixel. We propagate $\sigma_{\rm{rx+HI}}(x,y,v)$ to  estimate the corresponding contribution to the \HI\ column density image by adding noise velocity pixels in quadrature:     
\begin{equation}\label{eq:noise2}
        \sigma'_{\rm{rx+HI}} (x,y) = \sqrt{\Sigma_{v}({\sigma_{\rm{rx+HI}}(x,y,v)}^2)}
\end{equation}
    The above method produces an image $\sigma'_{\rm{rx+HI}} (x,y)$, where each spatial pixel shows the uncertainty value for the \HI column density resulting from the combined effects of the ALFA receiver, the \HI\ brightness temperature, and the scanning/calibration artifacts. Typical pixel values in this image are 1.3 to 3.3 K km s$^{-1}$ .

    We now calculate the uncertainty image associated purely with the \HI\ line, $\sigma_{\rm HI}$, as:    
\begin{equation}\label{eq:noise4}
        \sigma'_{\rm{rx+HI}} = \sqrt{\sigma_{\rm{rx}}^2 + \sigma_{\rm{HI}}^2}
\end{equation}
    and therefore:    
\begin{equation}\label{eq:noise5}
        \sigma_{\rm{HI}} = \sqrt{\sigma_{\rm{rx+HI}}^{'2} - \sigma_{\rm{rx}}^2}  .
\end{equation}
    Both $\sigma'_{\rm{rx+HI}}$ and $\sigma_{\rm{HI}}$
    are shown in Figure \ref{fig:off_HI}.
    To produce the $N_{\rm{on-HI}} (x,y)$ image (shown in Figure \ref{fig:off_HI}), 
    for each spatial pixel, a random number generator is used to draw a random number from a normal distribution with standard deviation equal to $\sigma_{\rm{HI}} (x,y)$.

    Once the non-\HI\ ($N_{\rm{off}}$) and \HI\ noise ($N_{\rm{on-HI}}$) images are combined, the final step in creating the noise image involves identifying and replacing any extreme outliers (hot pixels with values $\geq |5 \sigma|$). This step is necessary to eliminate extreme sharp discontinuities that can cause artificial effects on the SPS via Gibbs ringing.

\subsection{Constructing the noise SPS data cube from the noise image}\label{subsec:noiseSPS}

    Starting with Equation (\ref{eq:obs_sps}): 
\begin{equation}\label{eq:noise6}
    P_{\rm{obs}}(k)  = P_{\rm{beam}}(k) \times P_{\rm{model}} +  P_{\rm{noise}}(k) 
\end{equation}
    where $P_{\rm{noise}}(k)$ is the SPS of the noise image, $P_{\rm{beam}}(k)$ is the SPS of the ALFA's beam, and $P_{\rm{model}}$ is the SPS of the underlying \HI\ distribution unaffected by instrumental effects (noise, calibration), as explained in Section 3.2. 

    For the rolling SPS derivation, we derive the noise SPS for each three degree square sub-image from the full GALFA-HI noise image. To improve statistics for any give noise SPS, we parameterize $P_{\rm{noise}}(k)$ as a function of the standard deviation value of the input noise sub-image, $sd = \sigma(N_{\rm total})$ estimated for each 3-degree kernel, and write it as $P_{\rm{noise}}(k) = C \times Q(k,sd)$. Here, $C$ is a constant which varies as a function of spatial location and $Q(k,sd)$ is a function dependent on both $k$ and the standard deviation ($sd$) of the input noise sub-image. Below, we explain and justify the reason for this functional form.

\begin{figure*}
    \centering
    \subfloat{{\includegraphics[scale=0.6]{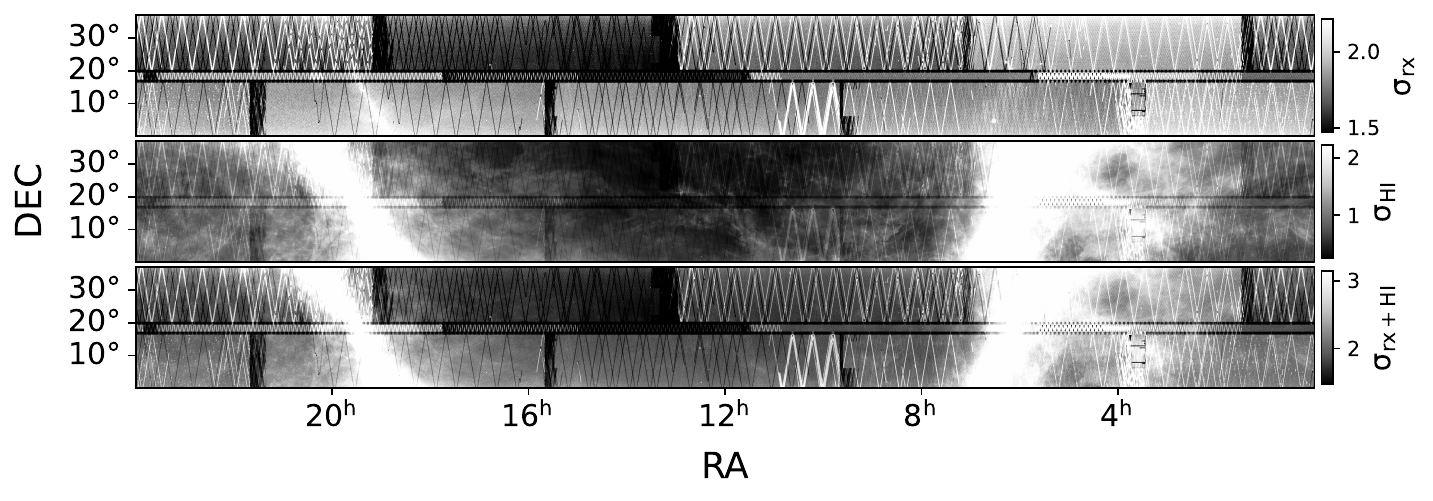}}}
    
    \subfloat{{\includegraphics[scale=0.6]{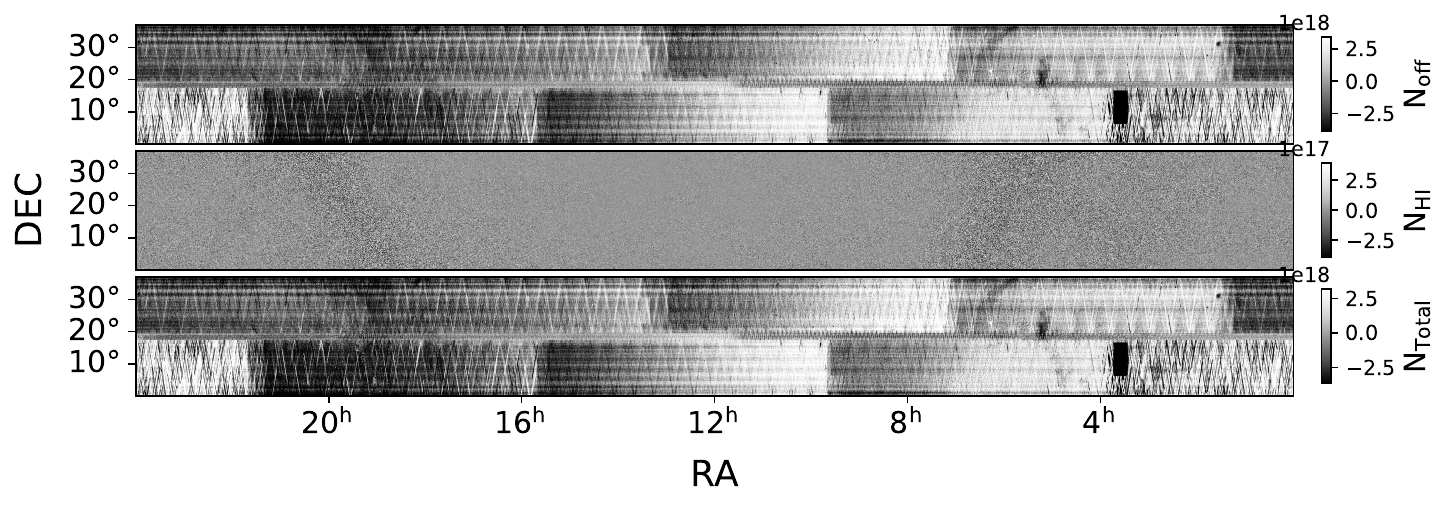}}}
    \caption{(Top 3 panels) 
    The standard deviation of the radiometer noise due to receiver temperature ($\sigma_{\rm rx}$), the standard deviation due to the \HI| brightness temperature ($\sigma_{\rm HI}$), and the standard deviation due to both receiver and HI brightness temperature ($\sigma_{\rm{rx+HI}}$).   
    (Bottom 3 panels) The noise contribution to the HI column density image: random fluctuations due to receiver temperature plus the systematic scanning uncertainty ($N_{\rm{off}}$), the contribution due to the brightness temperature that is velocity dependent ($N_{\rm{on-HI}}$), and the sum of these two contribution ($N_{\rm total}$).    }
    \label{fig:off_HI}
\end{figure*}

\begin{figure*}
    \centering
    \subfloat{\includegraphics[scale=0.4]{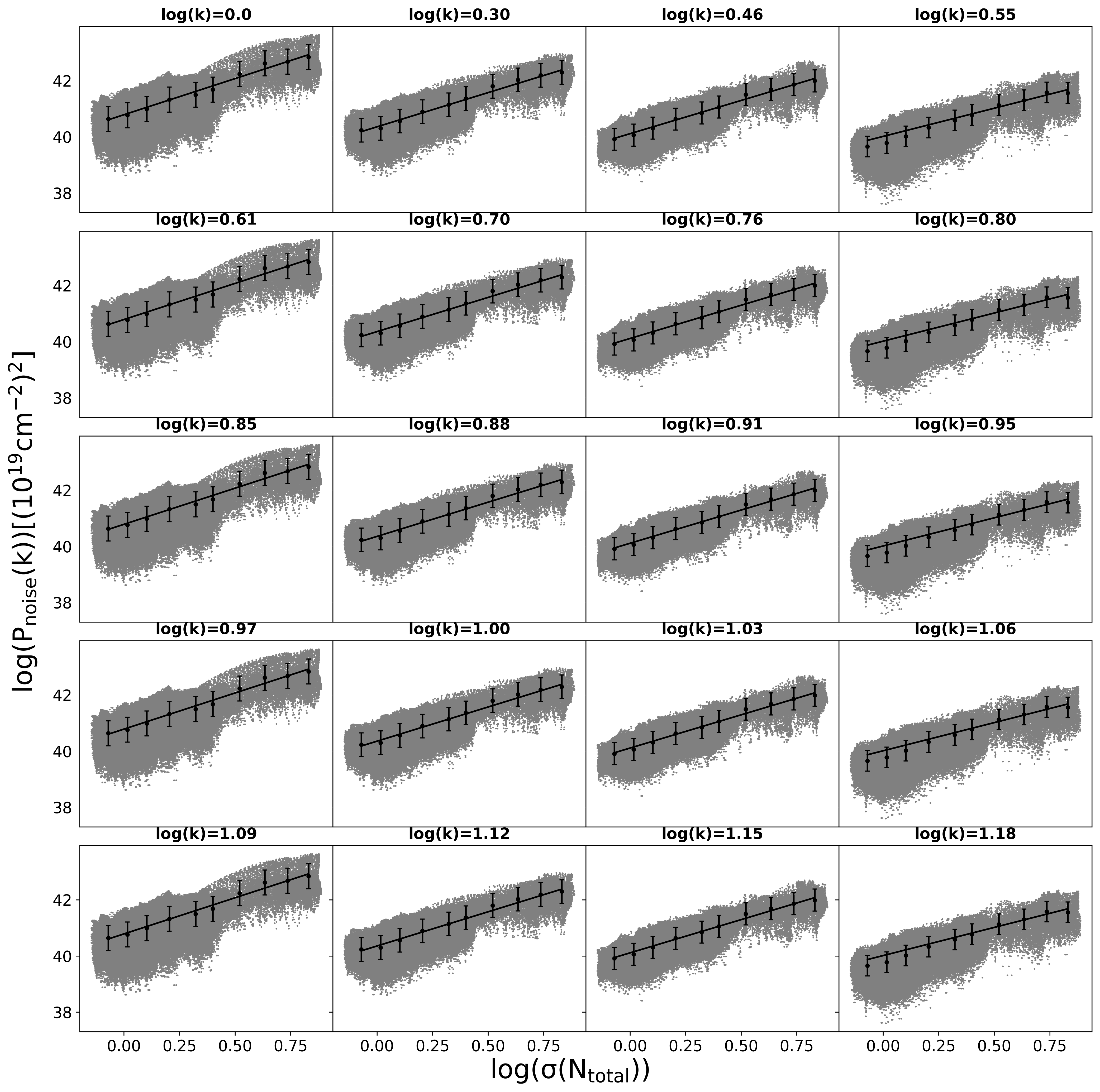}}
      \caption{The logarithmic power for the noise SPS, $P_{\rm noise}(k)$, as a function of the logarithm of $\sigma(N_{\rm total})$
       for the first 20 binned $k$ values obtained for all the 3-degree kernels are shown as grey data points. The mean values are overplotted in black. For each $k$, a linear function $Q(sd)$ was fitted to the data.}
    \label{fig:Q_noise}
\end{figure*}

\begin{figure*}
    \centering
    \subfloat{{\includegraphics[scale=1.0]{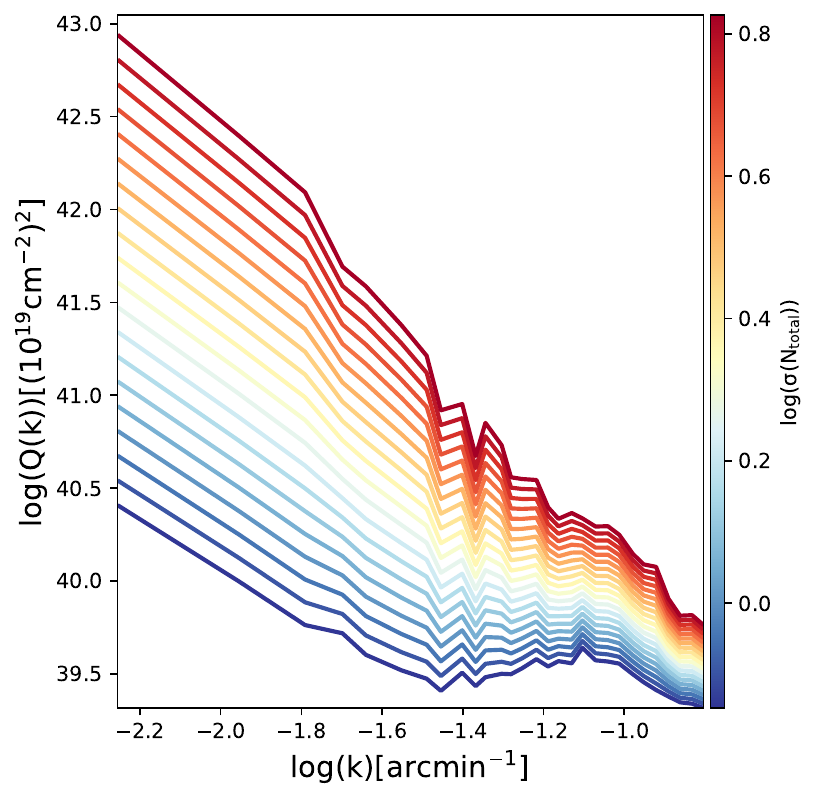}}}
   \caption{The parametrized noise SPS,  $Q(k, sd)$ for different values of $\sigma(N_{\rm total})$. 
}
    \label{fig:interval_sps}
\end{figure*}

    We first perform the rolling SPS analysis on the noise image $N_{\rm{total}}(x,y)$ to generate what we refer to as the raw noise SPS data cube. This cube contains the SPS of each 3-degree kernel of the the $N_{\rm{total}}(x,y)$ image. To enhance the statistics and smooth out these SPS functions before employing them in Equation (\ref{eq:noise6}), we compute an average noise SPS, denoted as $Q(k,sd)$.
    Due to the scanning and calibration artifacts, the primary source of noise in the GALFA-\HI\ data is the systematic noise (as seen in the bottom panel of Figure \ref{fig:off_HI}). Hence, the averaged noise SPS is significantly influenced by $\sigma_{\rm{rx}}$. Since each kernel covers a 3-degree area,
    $\sigma(N_{\rm{total}})$ is used as a proxy for $\sigma_{\rm{rx}}$. In Figure \ref{fig:Q_noise} we plot the $\log$ of  $\sigma(N_{\rm{total}})$ as a function of the $\log$ of $P_{\rm noise} (k)$ at each $\log(k)$ value used in the SPS calculation.
    This figure illustrates that for each $\log(k)$ value there is a linear relationship between 
     $\sigma(N_{\rm{total}})$ and $\log(P_{\rm noise})$.
    We therefore fit a linear function for each $\log(k)$ bin, providing the means for predicting the noise SPS of any noise sub-image by assessing  $\sigma(N_{\rm{total}})$. As a result, we have a family of about 100 $Q(k,sd)$ functions to represent $\log(P_{\rm noise})$. Figure \ref{fig:interval_sps} demonstrates the functional change of $\log(P_{\rm noise})$ as a function of $\sigma(N_{\rm{total}})$.

\begin{figure*}
    \centering
    \subfloat{{\includegraphics[scale=0.5]{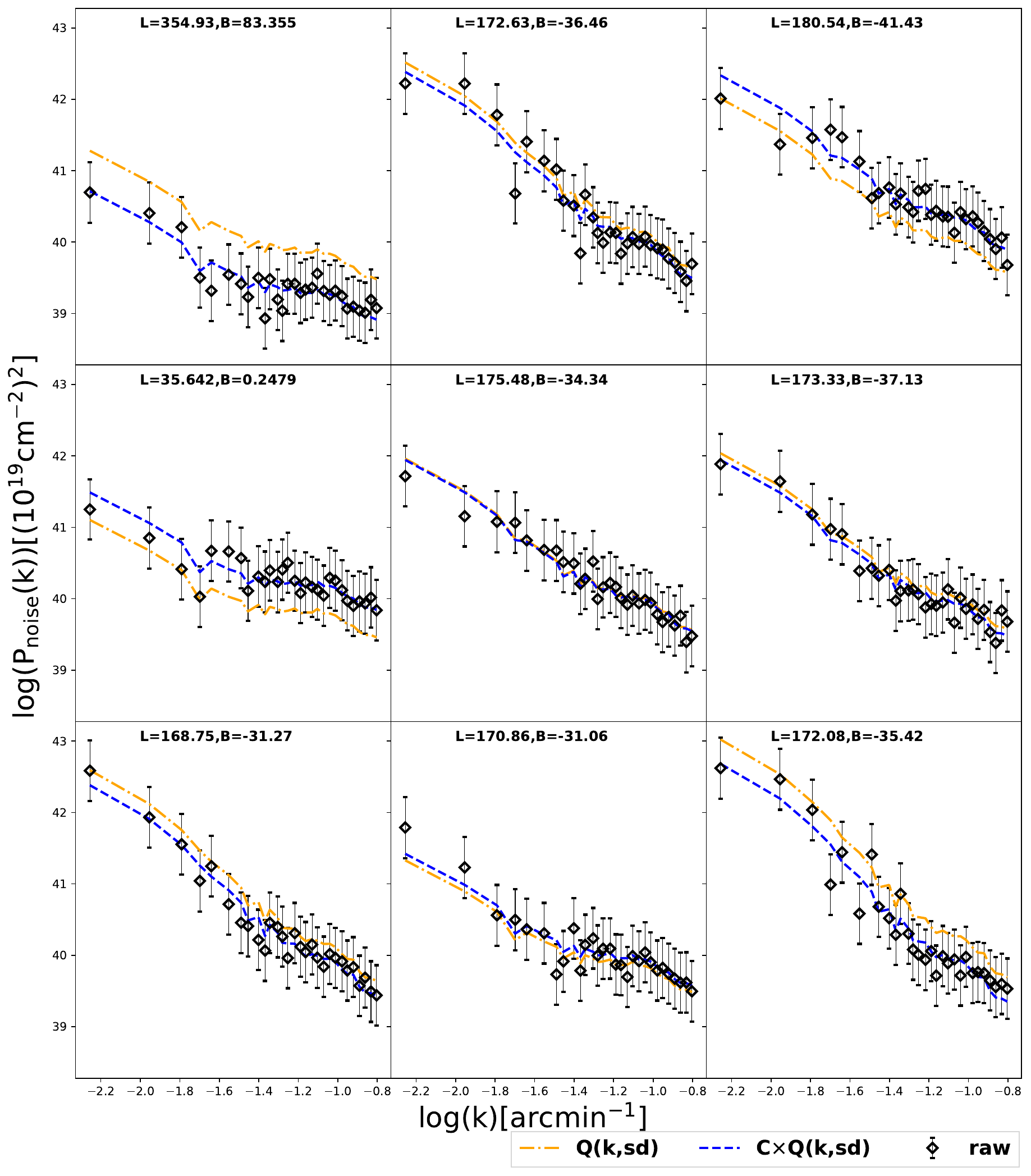}}}
    \caption{Examples of the noise SPS spectra for different kernels. The raw SPS data points, $P_{\rm noise}(k)$
    are shown as black data points. The parametrize noise SPS, $Q(k, sd)$ is shown in yellow, while the final parametrized noise SPS, 
    $C \times Q(k, sd)$, is shown in blue.  The SPS samples the range of angular scales from 3 degrees to 8 arcmin,
 which corresponds to spatial frequencies $k=5\times 10^{-3}$ to $10^{-1}$ arcmin$^{-1}$.}
    \label{fig:best_noise}
\end{figure*}

    Figure \ref{fig:best_noise} illustrates a sample of the raw noise SPS (curves black data points) over-plotted with the parameterized $Q(k,sd)$ noise power spectrum (orange line). As shown in Figure \ref{fig:off_HI}, 
    the total noise image has complex spatial structure and there are regions where the estimate of the local standard deviation will not provide a good fit to the noise SPS data. There are several such examples in Figure~\ref{fig:best_noise}, e.g. the top left panel ($(l, b)= (354.93, 83.355)$) where the orange line is clearly offset slightly from the data. To account for this, we estimate the constant value $C$ that is needed to shift the $Q(k,sd)$ noise power spectrum along the vertical axis and fit the data better. To estimate $C$ for each 3-degree kernel, we
    calculate the mean difference between $Q(k, sd)$ and $P_{\rm noise}(k)$ across different $k$ values. This simple procedure results in the $C\times Q(k,sd)$ noise power spectrum that is shown as a blue line in Figure \ref{fig:best_noise}.

\subsection{Final SPS after considering noise and beam effects}\label{subsec:corrrected_sps}

\begin{figure*}
    \subfloat{\includegraphics[width=15cm] {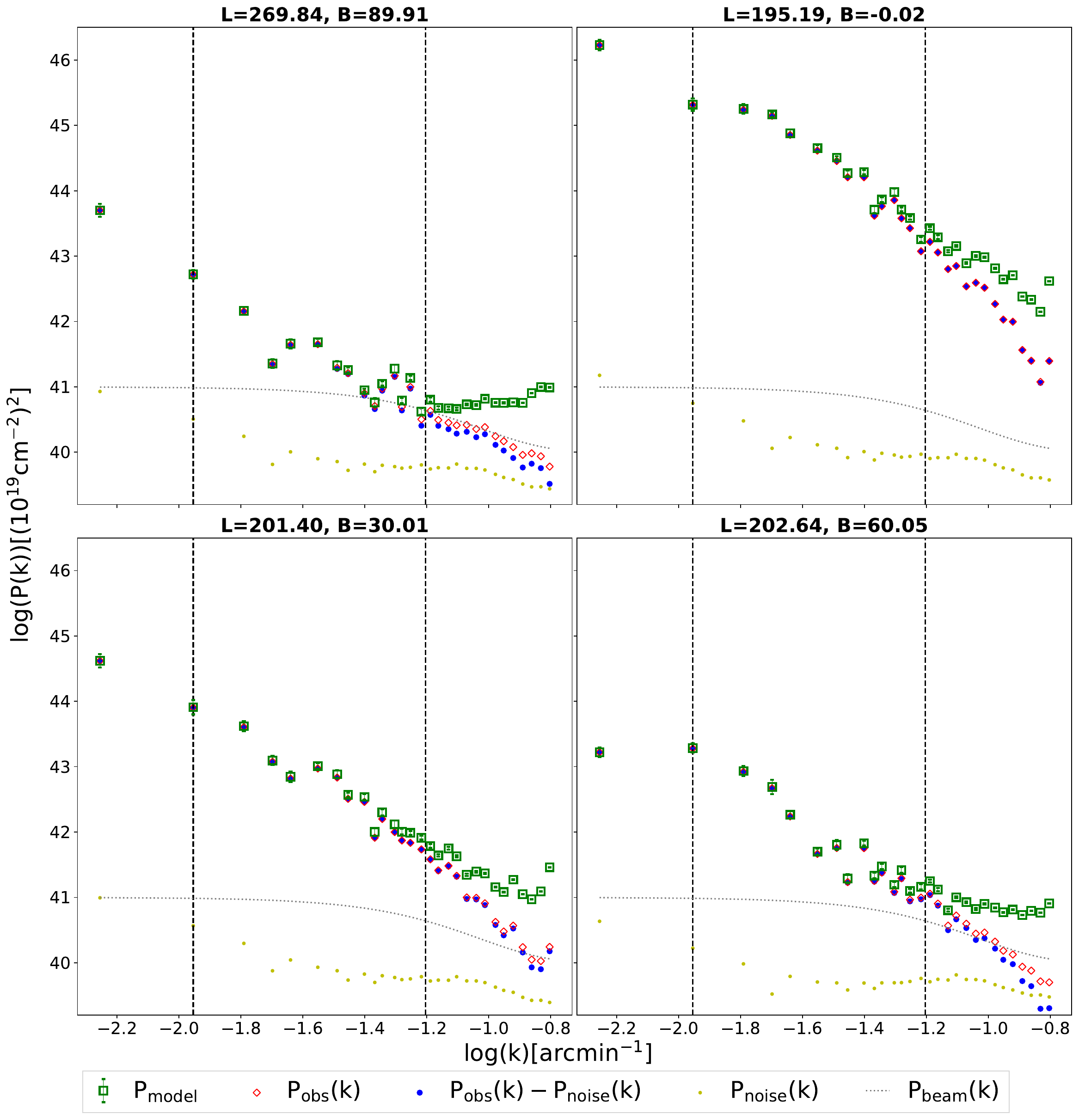}}
    \caption{Examples of the \HI\ SPS for four different regions with individual contributions from Equation (\ref{eq:obs_sps}). Within each panel, the grey, yellow, red, blue, and green data points represent the beam, total noise, observed, observed-beam, and final SPS, respectively. Furthermore, the dashed lines in the figure indicate the range of fitting for both large and small scales as 1.5\degree and 16\arcmin,  which corresponds to spatial frequencies $k=10^{-2}$ to $\sim 6 \times 10^{-2}$ arcmin$^{-1}$.}
    \label{fig:4POS_SPS}
\end{figure*}

    To demonstrate the entire process of accounting for the beam and noise effects, in Figure \ref{fig:4POS_SPS}, we present a sample of SPS curves calculated for 3-degree kernels at four distinct latitudes spanning from the Galactic plane to the Galactic pole. The observed SPS, $P_{\rm obs} (k)$ is shown with red data points in Figure \ref{fig:4POS_SPS}. The symmetrical Gaussian beam shape, $P_{\rm beam} (k)$ is shown as grey dashed curves. The noise SPS, $C \times Q(k, sd)$ is displayed as yellow data points in the same figure. Finally, $[P_{\rm obs} -  C \times Q(k, sd)]/P_{\rm beam}$ is shown as green data points. The range of $k$-values for fitting is restricted to 1.5-degree and 4 $\times$ FWHM for all kernels, as explained in the text.

\end{document}